\documentclass[runningheads]{svmult}

\usepackage{makeidx}   
\usepackage{graphicx}  
\usepackage{subeqnar}  
\usepackage{multicol}  
\usepackage{physprbb}  
\makeindex             

\newcommand{\arcsec}{\hbox{$^{\prime\prime}$}}
\begin{document}
\title*{VLT spectroscopy of NGC\,3115 globular clusters}
\toctitle{VLT spectroscopy of NGC\,3115 globular clusters}
%
%
\titlerunning{VLT spectroscopy of NGC\,3115 globular clusters}
%
\author{Harald Kuntschner\inst{1}
\and Bodo L. Ziegler\inst{2}\inst{,}\inst{3}
\and R.M. Sharples\inst{4}
\and Guy Worthey\inst{5}
\and Klaus J. Fricke\inst{3}}
\authorrunning{Harald Kuntschner et al.}
%
%
\institute{European Southern Observatory, Karl-Schwarzschild-Str. 2,
              85748 Garching bei M\"unchen, Germany
\and Academy of Sciences, Theaterstr. 7, 37073 G\"ottingen, Germany
\and Universit\"atssternwarte, Geismarlandstrasse 11, 37083 G\"ottingen, Germany
\and Department of Physics, University of Durham, Durham DH1 3LE, UK
\and Department of Physics, Washington State University, 1245 Webster Hall, Pullman, WA 99164-2814, USA
}

\maketitle              

\begin{abstract}
  We present results derived from VLT--FORS2 spectra of 17 globular
  clusters associated with the nearby lenticular galaxy NGC\,3115.
  Comparing line-strength indices to new stellar population models by
  Thomas et al. we determine ages, metallicities and element abundance
  ratios. Our data are also compared with the Lick/IDS observations of
  Milky Way and M\,31 globular clusters. Our best age estimates show
  that the observed clusters which sample the bimodal colour
  distribution of NGC\,3115 globular clusters are coeval within our
  observational errors (2--3 Gyr). Our best calibrated age/metallicity
  diagnostic diagram (H$\beta$ {\em vs}\/ [MgFe]) indicates an absolute
  age of 11--12 Gyr consistent with the luminosity weighted age for the
  central part of NGC\,3115. We confirm with our accurate line-strength
  measurements that the $(V-I)$ colour is a good metallicity indicator
  within the probed metallicity range ($-1.5 < \mathrm{[Fe/H]} < 0.0$).
  The abundance ratios for globular clusters in NGC\,3115 give an
  inhomogeneous picture. We find a range from solar to super-solar
  ratios for both blue and red clusters. This is similar to the data
  for M\,31 while the Milky Way seems to harbour clusters which are
  mainly consistent with $[\alpha / \mathrm{Fe}] \simeq 0.3$.
\end{abstract}

\section{The Sample and observations}
The candidate GCs were selected from the HST/WFPC2 investigation of
Kundu \& Whitmore (1998) who detected 144 globular cluster (herafter
GC) candidates in the central region of NGC\,3115. In order to keep
integration times reasonably short only clusters with $V<22$ (the peak
of the GC luminosity function is at $V=22.37\pm0.05$) were selected
while keeping a balance between red ($V-I < 1.06$) and blue ($V-I \ge
1.06$) clusters. The overall GC population
of NGC\,3115 shows a clear bimodal colour distribution with mean
metallicities at $\mathrm{[Fe/H]} \simeq -0.37$ and $\mathrm{[Fe/H]}
\simeq -1.36$ as estimated from the $V-I$ colours.

In order to utilize the full field-of-view of FORS2 we supplemented
this list with GC candidates from a low-resolution spectroscopy survey
reported in Kavelaars (1998) and also placed some slits on promising
objects without prior information. In total 29 spectra were obtained.
Applying a cut in recession velocity and S/N (S/N~$\ge 12$ per pixel) 
yielded a final sample of 17 GCs.

The observations were carried out with FORS2 on the VLT with the blue
600 l/mm grism and 1\arcsec\/ wide MXU slitlets giving a resolution of
$\sim$5~\AA\/ (FWHM). The total exposure time was 12\,440~s.

\section{Abundance Ratios}
\begin{figure}[htb]
\begin{center}
\includegraphics[width=1.0\textwidth]{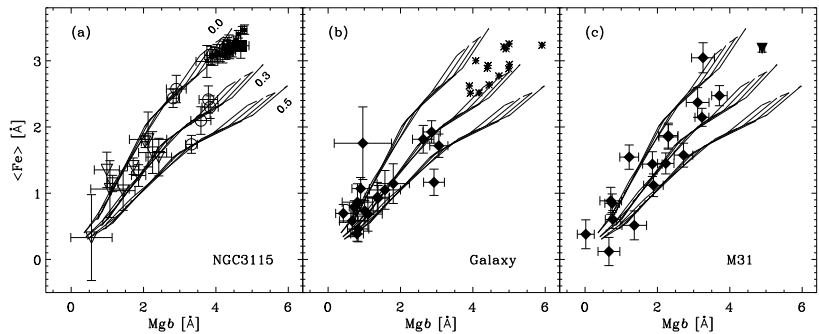}
\end{center}
\caption[]{Probing the [Mg/Fe] ratios of globular
  clusters in a Mg\,$b$\/ {\em vs}\/ $\langle$Fe$\rangle$ diagram.
  (\textbf{a}) Our sample of globular clusters in NGC\,3115 is shown as
  triangles (blue clusters) and circles (red clusters). The large
  filled square represents the centre of NGC\,3115 taken from Trager et
  al. (1998) and the small filled squares represent the data of Fisher,
  Franx \& Illingworth (1996) which cover radii up to 40\arcsec\/ along
  the major axis. (\textbf{b}) Milky Way globular clusters observed
  with the Lick/IDS instrumentation. (\textbf{c}) Globular clusters in
  M\,31 observed with the Lick/IDS instrumentation. The filled triangle
  represents the centre of M\,31 taken from Trager et al. (1998).
  Overplotted in all panels are models by Thomas et al. (2002) with
  abundance ratios of [Mg/Fe]=0.0., 0.3, 0.5 as indicated in the left
  panel. The models span a range in age (3--12~Gyr) and metallicity
  ($\mathrm{[Fe/H]} = -2.35~\mathrm{to}~ +0.3$)}
\label{fig:ratio}
\end{figure}

Theory of chemical enrichment predicts that stellar populations created
in a short burst of star formation show elevated magnesium-to-iron
abundance ratios while extended periods of star formation lead to
roughly solar abundance ratios. Abundance ratios for integrated stellar
populations can be best explored in a Mg\,$b$ {\em vs}\/
$\langle\mathrm{Fe}\rangle$ index diagram ($\langle\mathrm{Fe}\rangle =
(\mathrm{Fe5270} + \mathrm{Fe5335})/2$). Figure~\ref{fig:ratio} shows
this diagnostic line-strength index diagram for GCs in NGC\,3115, the
Milky Way and M\,31. We find that GCs in NGC\,3115 show a range in
abundance ratios similar to M\,31 from [Mg/Fe] $=0.0$ to 0.5 by comparing 
with models of Thomas et al. (2002). This applies to both the red and blue
GCs in NGC\,3115.

\section{Age and metallicity estimates}
\begin{figure}[htb]
\begin{center}
\includegraphics[width=1.0\textwidth]{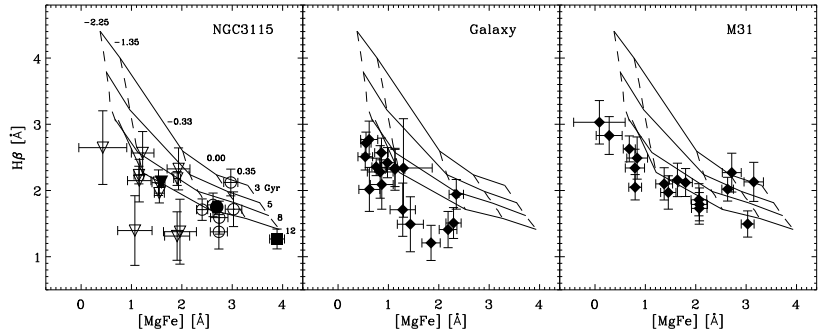}
\end{center}
\caption[]{Age and metallicity diagnostic diagrams using as metallicity
  indicator [MgFe] ($\mathrm{[MgFe]} = \sqrt{\mathrm{Mg}\,b \times
    \langle \mathrm{Fe} \rangle}$) and as age indicator H$\beta$.
  H$\beta$ and [MgFe] are not significantly influenced by abundance
  ratios.  (\textbf{a}) Our sample of globular clusters is shown as
  triangles (blue clusters) and circles (red clusters). The error
  weighted means of the blue and red clusters are shown as filled
  symbols. The filled black square represents the centre of NGC\,3115.
  (\textbf{b}) and (\textbf{c}) The Lick/IDS data for Milky Way and
  M\,31 globular clusters are shown. Overplotted are solar--abundance
  SSP models by Thomas et al. (2002) for metallicities $\mathrm{[Fe/H]}
  = -2.25, -1.35, -0.33, 0.00, 0.35$ (dashed lines left to right)
  and ages 3, 5, 8, and 12~Gyr (solid lines top to bottom).}
\label{fig:age}
\end{figure}

\begin{figure}[b]
\begin{center}
\includegraphics[width=1.0\textwidth]{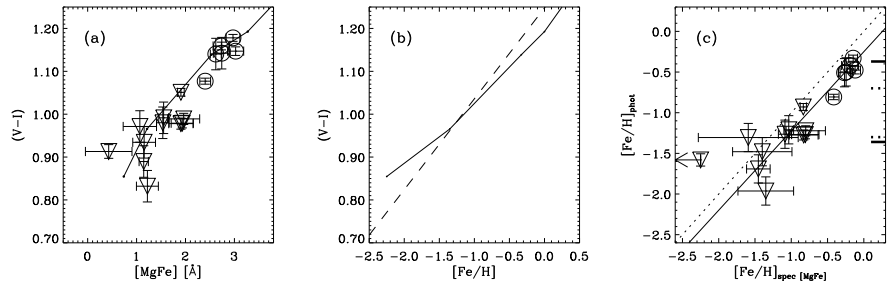}
\end{center}
\caption[]{Comparison of photometric and spectroscopic metallicity
  estimates. (\textbf{a}) shows a direct comparison of observed
  quantities. In (\textbf{b}) we show a comparison between an empirical
  calibration of $(V-I)$ colour against metallicity (dashed line)
  and predictions from stellar population models (solid line).
  (\textbf{c}) Photometric metallicities are estimated from the $(V-I)$
  colour, while spectroscopic metallicities are derived from the [MgFe]
  index assuming a constant age of 12~Gyr. The dotted line is the
  line of unity while the solid line is a linear fit to the data.}
\label{fig:comp}
\end{figure}

In Figure~\ref{fig:age} we show age/metallicity diagnostic diagrams using
the H$\beta$ Balmer absorption and the mean metallicity ([Fe/H]) indicator
[MgFe]. The error weighted mean ages of the blue and red GC populations are
very similar to within 2--3~Gyr and favour absolute model ages of 11--12\,Gyr.
There is a clear correlation between
the $(V-I)$ colour and the metallicity as measured
from metal absorption lines (see also Figure~\ref{fig:comp}).

\section{Conclusions}
\begin{itemize}

\item The GCs in NGC\,3115 show a range of abundance ratios as
  estimated by the strength of Mg and Fe lines. Specifically we find
  for both red and blue clusters solar as well as super-solar values
  (up to $\mathrm{[Mg/Fe]} \simeq +0.5$). Lick/IDS data of M\,31 GCs
  show a similar distribution while Milky Way GCs are consistent with a
  constant value of $\mathrm{[Mg/Fe]} \simeq 0.3$. 
  
\item Our analysis of the H$\beta$ {\em vs}\/ [MgFe] age/metallicity
  diagnostic diagrams shows that the red and blue GC populations are
  coeval within $\approx$2~Gyr. Assuming a correct calibration of the
  model line-strengths we estimate a mean age of $11-12$~Gyr.
  
\item We present a comparison of photometric and spectroscopic
  metallicity determinations and find a good linear relation in the
  metallicity range probed by our sample of NGC\,3115 clusters ($-1.5 <
  \mathrm{[Fe/H]} < 0.0$). The photometric estimates are systematically
  lower ($\simeq -0.26$) in comparison with our spectroscopic
  measurements.
  
\item The existence of solar as well as elevated Mg-to-Fe ratios at a
  given metallicity for GCs in NGC\,3115 indicates that a simple
  scenario of two distinct star-formation episodes is not sufficient to
  explain the formation of this galaxy. Probably a realistic model
  needs to incorporate more than two distinct star-formation events.

\end{itemize}

For a detailed presentation of this work we refer the reader to our paper
Kuntschner et al., A\&A acc. (astro-ph/0209129).

\subsubsection{Acknowledgements:}
Based on observations collected by Dr. J. Heidt (Heidelberg) at the European 
Southern Observatory, Cerro Paranal, Chile (ESO No. 66.B-0131).
Part of this work was supported by the Volkswagen Foundation
(I/76\,520).

%

\end{document}